# Agile Modeling with the UML

Bernhard Rumpe
Software & Systems Engineering,
Technische Universität München
85748 Munich/Garching, Germany,

This paper discusses a model-based approach to software development. It argues that an approach using models as central development artifact needs to be added to the portfolio of software engineering techniques, to further increase efficiency and flexibility of the development as well as quality and reusability of the results. Two major and strongly related techniques are identified and discussed: Test case modeling and an evolutionary approach to model transformation.

## 1 Portfolio of Software Engineering Techniques

Software has become a vital part of our lives. Embedded forms of software are part of almost any technical device from coffee machine to cars, the average household uses several computers, and the internet and telecommunication world has considerably changed our lives. All these machines are driven by a huge variety of software. Software that must never fail, must be updated dynamically, must continuously evolve to meet customers needs, must contain its own diagnosis and "healing" subsystems, etc. Software is used to for a variety of jobs, starting with the control of many kinds of processes up to simply to entertain. Software can be as small as a simple script or as complex as an entire operating or enterprise resource planning system.

Nowadays, there is some evidence that there will not be a single notation or process that can cover the diversity of today's development projects. Projects are too different in their application domain, size, need for reliability, time-to-market pressure, and the skills and demands of the project participants. Even the UML [1], which is regarded as a de-facto standard, is seen as a family of languages rather than a single notation and by far doesn't cover all needs. This leads to an ongoing proliferation of methods, notations, principles, techniques and tools in the software engineering domain. Some indicators of this ongoing diversity are:

- New programming languages, such as Python [2] without strong typing systems, but powerful capabilities for string manipulation and dynamic adaptation of their own program structure compete with Java, C++ and other conventional languages.

- The toolsets around XML-based standards [3] are widely and successfully used, even though they basically reinvent all compiler techniques known for 20 years.
- Methods like Extreme Programming [4] or Agile Software Development [5] discourage the long well known distinction between design and implementation activities and abandon all documentation activities in favor of rigorous test suites.
- Upcoming CASE tools allow to generate increasing amounts of code from UML models, thus supporting the OMG's initiative on "Model Driven Architecture" (MDA) [6]. MDA's primary purpose is to decouple platform-independent models and from platform-specific, technical information. This should increase the reusability of both.
- Completely new fields, like security, need new foundations embedded in practical tools. New logics and modeling techniques [7] are developed and for example used to verify protocols between mutually untrusted partners.

From this observations it is evident that in the foreseeable future we will have a *portfolio of software engineering techniques* that enables developers and managers to select appropriate processes and tools for their projects. To be able for such a selection developers need to be aware of this *portfolio of software engineering techniques* and master at least a comprehensible subset of these techniques. Today, however, it is not clear which elements the portfolio should have, how they relate, when they are applicable, and what their benefits and drawbacks are. The software and systems engineering community therefore must reconsider and extend its portfolio of software engineering techniques incorporating new ideas and concepts, but also try to scientifically assess the benefits and limits of new approaches. For example:

- Lightweight projects that don't produce requirement and design documentation need intensive communications and can hardly be split into independent subprojects. Thus they don't scale up to large projects. But where are the limits? A guess is, around 10 people, but there have been larger projects reportedly "successful" [23].
- Formal methods have built a large body of knowledge, but how can this knowledge successfully and goal-oriented be applied in today's projects? A guess seems to be, formal methods spread best, if embodied in practical tools, using practical and well known notations.
- Product reliability need not be 100% for all developments and already in the first iteration. But how to predict reliability from project metrics and how to adapt the project to increase reliability and accuracy to the desired level while minimizing the project/product costs?
- Promising techniques such as functional programming still need to find their place in practical software engineering, where they don't play an appropriate role today. A guess is that e.g., functional programming will be combined with object-techniques such that functional parts can be embedded in conventional programs.

Based on the observation for a general demand for a broad portfolio of SE techniques, we will in the following examine two trends that currently and in the foreseeable future will influence software engineering. These are on the one hand, the modeling notation UML and on the other hand agile development techniques. Although, these two trends currently work against each other, we can see that there is potential for their combination that takes benefits from both.

Section 2 discusses synergies and problems of using models for a variety of activities, including programming. Section 3 explores the possibilities of increasing efficiency and reducing development process overhead that emerges from the use of models as code and test case descriptions. In Section 4 the overall scenario of a model based test approach is discussed. Section 5 finally presents the benefits of an evolutionary approach to modeling in combination with an intensive, model-based test approach. In particular, the usability of tests as invariant observations for model-transformations is explored. For sake of conceptual discussions, technical details are usually omitted, but can be found in [8].

## 2 Modeling meets Programming

UML [1] undoubtedly has become the most popular modeling language for software intensive systems used today. Models can be used for quite a variety of purposes. Among them are most common:

- Informal sketches are used for *communication*. Such a sketch is usually drawn on paper and posted on a wall, but not even used for documentation.
- More precisely defined and larger sets of diagrams are used for *documentation* of requirements and design. But requirements are usually captured in natural language and a few top-level and quite informal drawings that denote an abstract architecture, use cases or activity diagrams.
- Architecture and designs are captured and documented with models. In practice, these models are used for *code generation* increasingly often.

More sophisticated and therefore less widespread uses of models are analysis of certain features (such as throughput, failure likelihood) or development of tests from models. Many UML-based tools today offer functionality to directly simulate models or generate at least parts of the code. As tool vendors work hard on continuous improvement of this feature, this means a sublanguage of UML will become a high-level programming language and modeling at this level becomes identical to programming. This raises a number of interesting questions:

- Is it critical for a modeling language to be also used as programming language? For example analysis and design models may become overloaded with details that are not of interest yet, because modelers are addicted to executability.

- Is the UML expressive enough to describe systems completely or will it be accompanied by conventional languages? How well are these integrated?
- How will the toolset of the future look like and how will it overcome round trip engineering (i.e. mapping code and diagrams in both directions)?
- What implications does an executable UML have on the development process?

In [8,9] we have discussed these issues and have demonstrated, how the UML in combination with Java may be used as a high-level programming language. But, UML cannot only be used for modeling the application, but more importantly for modeling tests on various levels (class, integration, and system tests) as well. Executable models are usually less abstract than design models, but they are more compact and abstract as the implementation.

One advantage of using models for test case description is that application specific parts are modeled with UML-diagrams and technical issues, such as connection to frameworks, error handling, persistence, or communication are handled by the parameterized code generator. This basically allows us to develop models independent of any technology or platform, as for example proposed in [10]. Only in the generation process platform dependent elements are added. When the technology changes, we only need to update the generator, but the application defining models can directly be reused. This concept also directly supports the above mentioned MDA-Approach [6] of the OMG. Another important advantage is that both, the production code and automatically executable tests at any level, are modeled by the same UML diagrams. Therefore developers use a single homogeneous language to describe implementation and tests. This will enhance the availability of tests already at the beginning of the coding activities. Similar to the "test first approach" [11,12], sequence diagrams are used for test cases and can be taken from the previously modeled requirements.

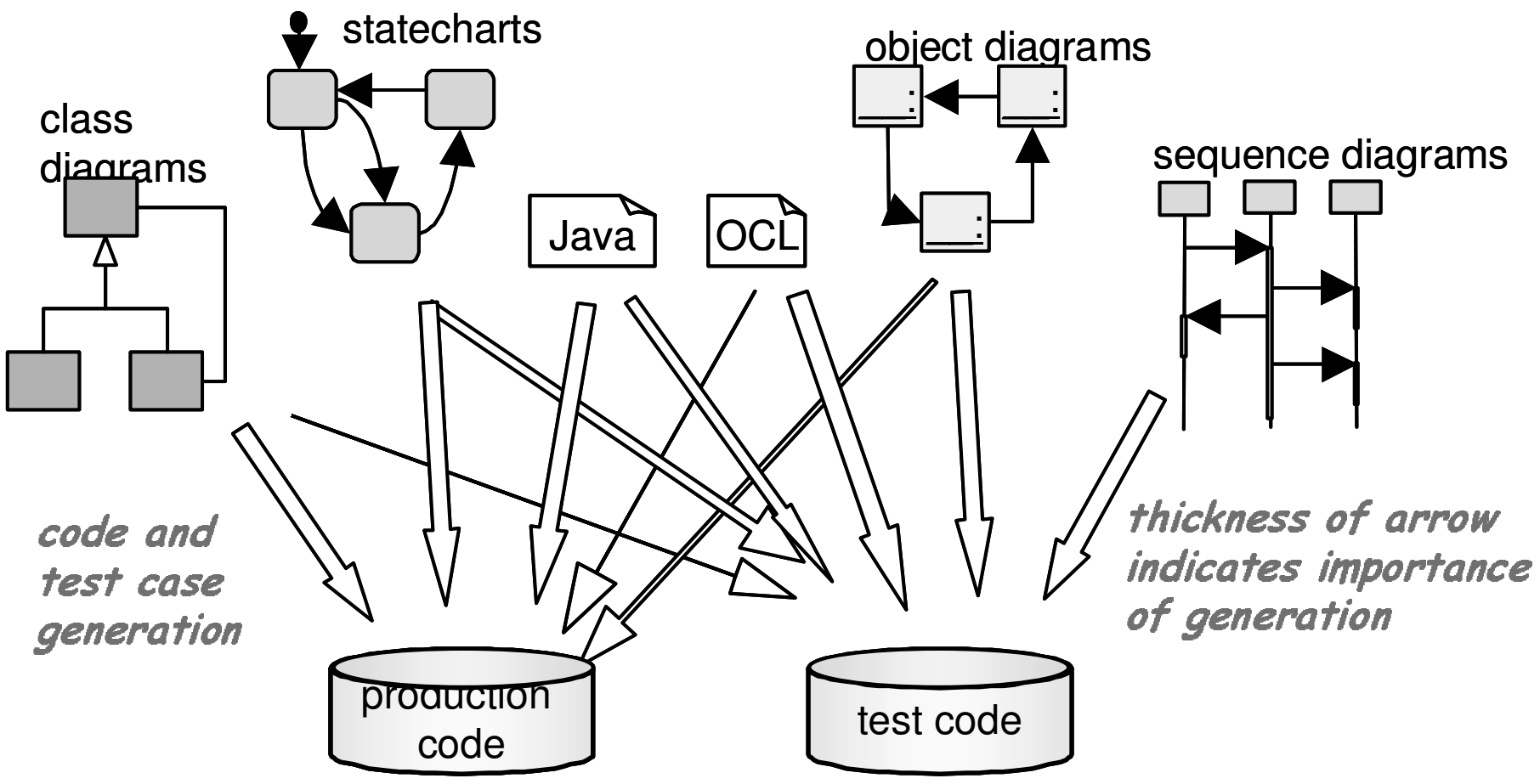

**Fig. 1.** Mapping of UML-models to code and test code.

Part of the UML-models (mainly class diagrams and statecharts) are used constructively, others are used for test case definition (mainly OCL, sequence and enhanced object diagrams). Fig. 1 illustrates the key mappings.

## 3 Agile Modeling: Using Models in Agile Projects

In the last few years a number of agile methods have been defined that share a certain kind of characteristics, described in [13]. Among these Extreme Programming (XP) [4] is the most widely used and discussed method. Some of the XP characteristics are:

- It focuses on the primary goal, the production code. Documentation instead is widely disregarded, but coding standards are enforced to document the code well.
- Automated tests are used on all levels. Practical experience shows, that when this is properly done, the defect rate is considerably low. Furthermore, the automation allows to repeat tests continuously.
- Very small iterations with continuous integration are enforced and the system is kept as simple as possible.
- Refactoring is used to improve the code structure and tests ensure a low defect rate introduced through refactoring.

The abandoning of documentation is motivated by the gained reduction of workload and the observation, that developers don't trust documents, because these usually are out of date. So, XP directly focuses on code. All design activities directly manifest in the code. Quality is ensured through strong emphasis on testing activities, ideally on development of the tests before the production code ("test first approach" [11]). An explicit architectural design phase is abandoned and the architecture emerges during coding. Architectural shortcomings are resolved through the application of refactoring techniques [14,15]. These are transformational techniques to refactor a system in small steps to enhance its structure. The concept isn't new [16], but through availability of tools and its embedding in XP, transformational development now becomes widely used.

When using an executable version of UML to develop the system within an agile approach, the development project should become even more efficient. On the one hand, through the abstractness of the platform independent models, these models are more compact and can more easily be written, read and understood than code. On the other hand in classic development projects these models are developed anyway. But, increased reuse of these models for later stages now becomes feasible through better assistance. Therefore, model-based development as proposed by the MDA-approach [6] becomes applicable. The separation of application models and platform specific parts that are combined through code generation only exhibits some characteristics of aspect oriented programming [17]. These UML-models also serve as up-to-date documentation much better than commented code does.

To summarize, Fig. 2 shows the techniques used on models. This is quite in contrast to [18], where models are only used as informal drawings on the wall without further impact.

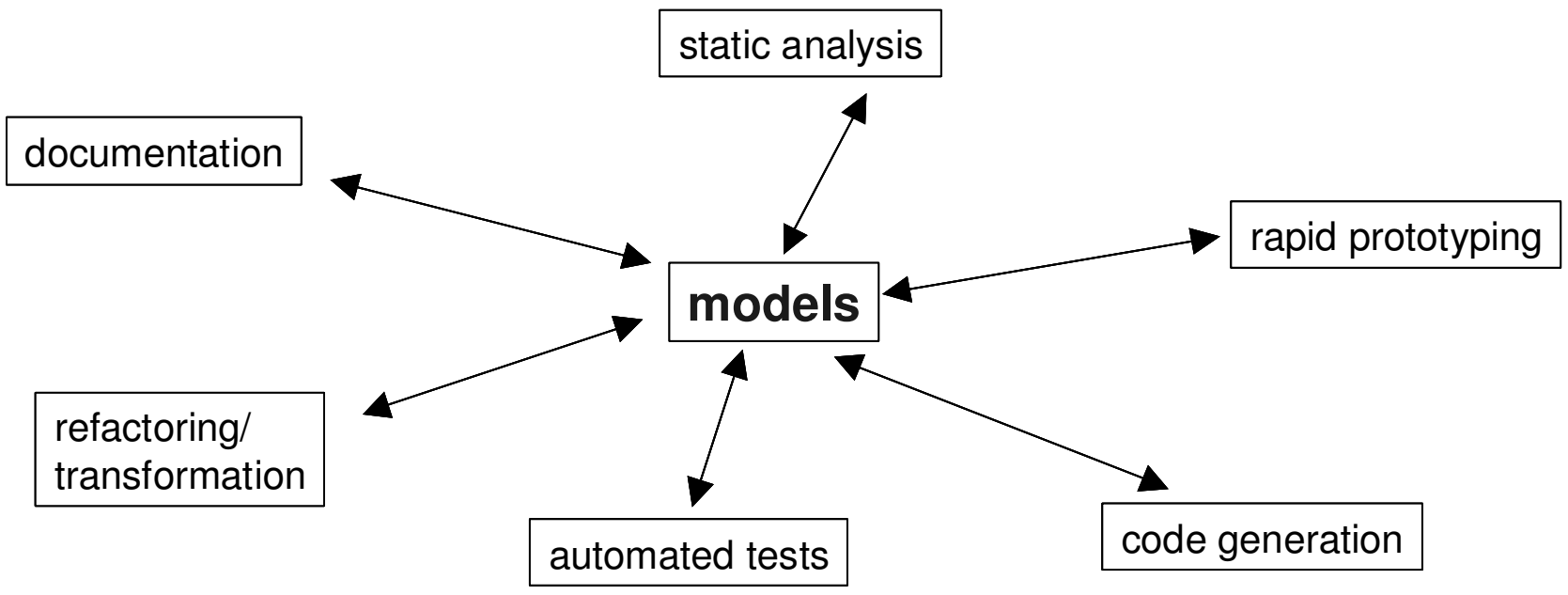

**Fig. 2.** The potential uses of UML-models.

## 4 Model-based Testing

There exists a huge variety of testing strategies [19,20]. The use of models for the definition of tests and production code can be manifold:

- Code or at least code frames can be generated from a design model.
- Test cases can be derived from an analysis or design model that is not used/usable for constructive generation of production code. For example behavioral models, such as statecharts, can be used to derive test cases that cover states, transitions or even paths.
- The modeling technique itself can be used to describe a test case or at least a part thereof.

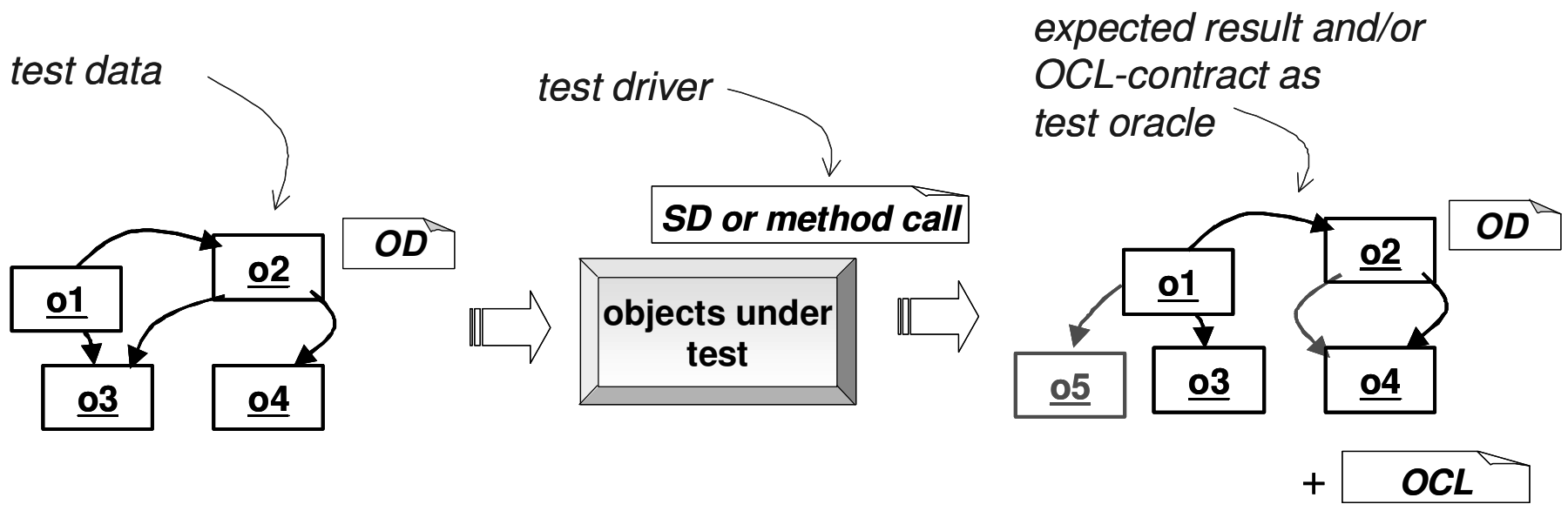

**Fig. 3.** Structure of a test modeled with object diagrams (OD), sequence diagram (SD) and the Object Constraint Language (OCL).

The first two uses are already discussed e.g. in [20]. Therefore, in this section we concentrate on the development of models that describe tests. A typical test, as shown in Fig. 3 consists of a description of the test data, the test driver and an oracle characterizing the desired test result. In object-oriented environments, the test data can usually be described by an object diagram (OD). It shows the necessary objects as well as concrete values for their attributes and the linking structure. The test driver can be modeled using a simple method call or, if more complex, a sequence diagram (SD). An SD has the considerable advantage that not only the triggering method calls can be described, but it is possible to model desired interactions and check object states during the test run.

For this purpose, the Object Constraint Language (OCL, [21]) is used. In the sequence diagram in Fig. 4, an OCL constraint at the bottom ensures that the new closing time of the auction is set to the time when the bid was submitted (bid.time) plus the extension time to allow competitors to react (the auction system containing this structure is in part described in [8,22]). Furthermore, it has proven efficient to model the test oracle using a combination of an object diagram and OCL properties. The object diagram in this case serves as a property description and can therefore be rather incomplete, just focusing on the desired effects. The OCL constraints used can also be general invariants or specific property descriptions.

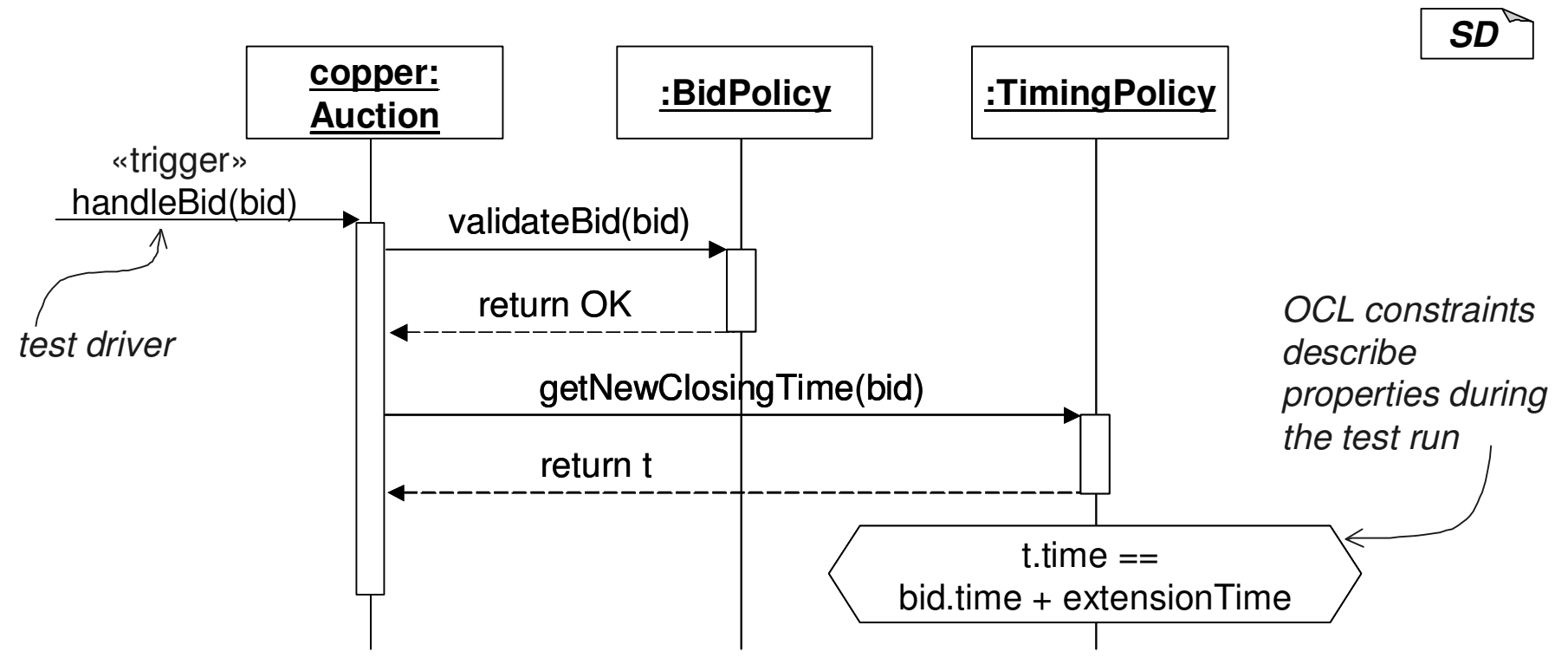

**Fig. 4.** A sequence diagram (SD) describing the trigger of a test driver and some test interactions as well as an OCL property that holds at that point of time.

As already mentioned, being able to use the same, coherent language to model the production system and the tests allows for a good integration between both tasks. It allows the developer to immediately define tests for the constructive model developed. It is imaginable that in a kind of “test-first modeling approach” the test data in form of possible object structures is developed before the actual implementation.

## 5 Model Evolution using Automated Tests

Neither code nor models are initially correct. For code, many sources of incorrectness can rather easily be analyzed using type checkers of compilers and automated tests that run on the code. For models this is usually a problem that leaves many errors undetected in analysis and design models. This is particularly critical as conceptual errors in these models are rather expensive if detected only late in the development process. The use of code generation and automated tests helps to identify errors in these models.

Besides detecting errors, which might even result from considerable architectural flaws, nowadays, it is expected that the development and maintenance process is capable of being flexible enough to dynamically react on changing requirements. In particular, enhanced business logic or additional functionality should be added rapidly to existing systems, without necessarily undergo a major re-development or re-engineering phase. This can be achieved at best, if techniques are available that systematically evolve the system using transformations. To make such an approach manageable, the refactoring techniques for Java [14] have proven that a comprehensible set of small and systematically applicable transformation rules seems optimal. Transformations, however, cannot only be applied to code, but to any kind of model. A number of possible applications are discussed in [16].

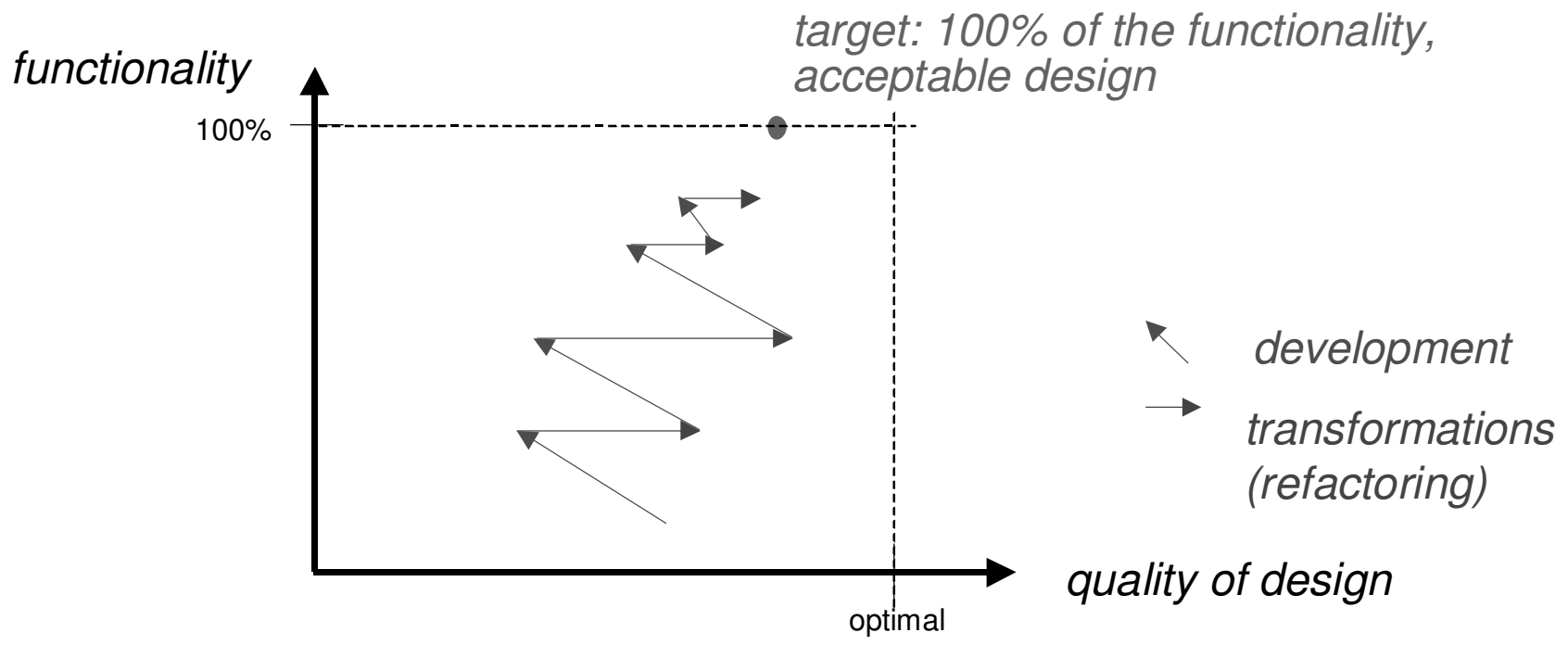

**Fig. 5.** Transformations to improve the quality of design opposed to development steps that add functionality.

Having a comprehensible set of model transformations at hand, model evolution becomes a crucial step in software development and maintenance. Architectural and design flaws can then be more easily corrected, superfluous functionality and structure removed, structure for additional functionality or behavioral optimizations be adapted, because models are more abstract, exhibit higher-level architectural and design information in a better way. During development, the situation can roughly be described with Fig. 5. It shows the dimension of functionality (measured for example in function points) and the “quality of design” (without a good metrics and therefore as informal concept). The development process tries to reach the 100% functionality while at the same time targets a reasonable good design.

The core development process ideally consists of step from two categories:

- Development (or programming) steps add functionality. But usually they in practice also introduce “erosion” of the design quality. For example repeated adding of new methods to a class overloads that class, simplifications of the code might come up, etc. Thus design quality usually suffers.
- Transformational (refactoring) steps build the second category. They improve structure and design, without changing the “externally observable behavior”.

Two simple transformation rules on a class diagram are shown in Fig. 6. The figure shows two steps that move a method and an attribute upward in the inheritance hierarchy. The upward move of the attribute is accompanied by the only context condition, that the other class “Guest” didn’t have an attribute with the same name yet. In contrast, moving the method may be more involved. In particular, if both existing method bodies are different, there are several possibilities: (1) Move up one method implementation and have it overridden in the other class. (2) Just add the method as abstract signature in the superclass. (3) Adapt the method implementations in such a way that common parts can be moved upward. This can for example be achieved by factoring differences between the two implementations of “checkPasswd” into smaller methods, such that at the end a common method body for “checkPasswd” remains. As a context condition, the moved method may not use attributes that are available in the subclasses only.

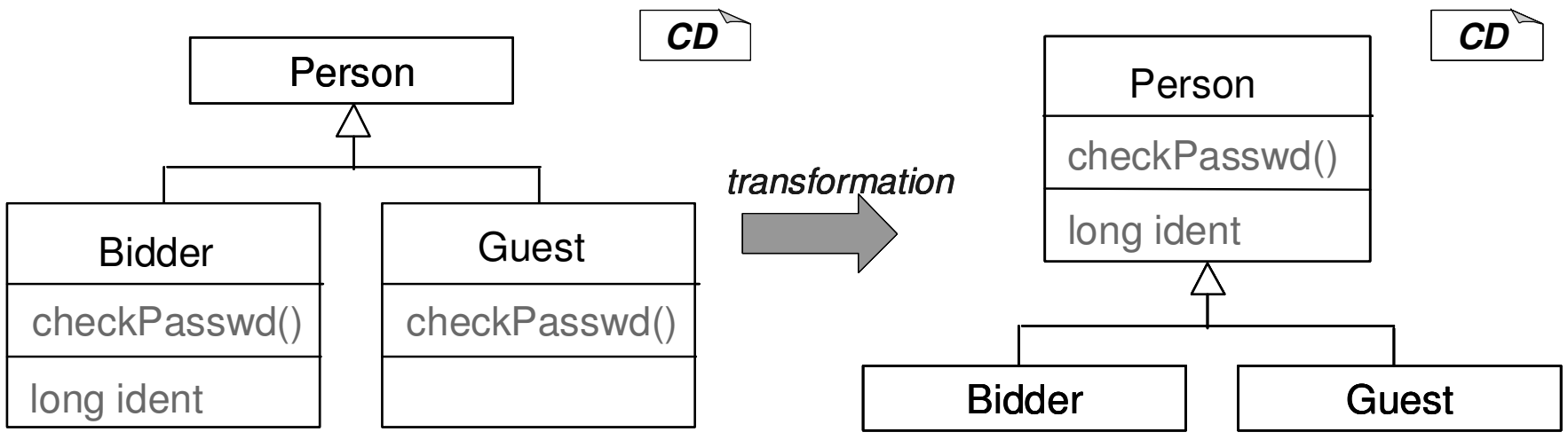

**Fig. 6.** Two transformational steps moving an attribute and a method along the hierarchy.

Many of the necessary transformation steps are as simple as the upward move of an attribute. However, others are more involved and their application comes with a larger set of context conditions and accompanying steps similar to the adaptation necessary for the “checkPasswd” method. These of course need automated assistance. The power of these simple and manageable transformation steps comes from the possibility to combine them and evolve complex designs in a systematic and traceable way.

Following the definition on refactoring [14], we use transformational steps for structure enhancement that does not affect “externally visible behavior”. For example both transformations shown in Fig. 6 do not affect the external behavior if made properly.

By "externally visible behavior" Fowler in [14] basically refers to behavioral changes visible to the user. This can be generalized by introducing an abstract "system border". This border serves as interface to the user, but may also act as interface to other systems. Furthermore, in a hierarchically structured system, we may enforce behavioral equivalence for "subsystem borders" already. It is therefore necessary to explicitly describe, which kind of behavior is regarded as externally visible. For this purpose tests are the appropriate technique to describe behavior, because (1) tests are already available through the development process and (2) tests are automated which allows us to check the effect of a transformation through inexpensive, automated regression testing.

A test case thus acts as an "observer" of the behavior of a system under a certain condition. This condition is also described by the test case, namely through the setup, the test driver and the observations made by the test. Fig.7 illustrates this situation.

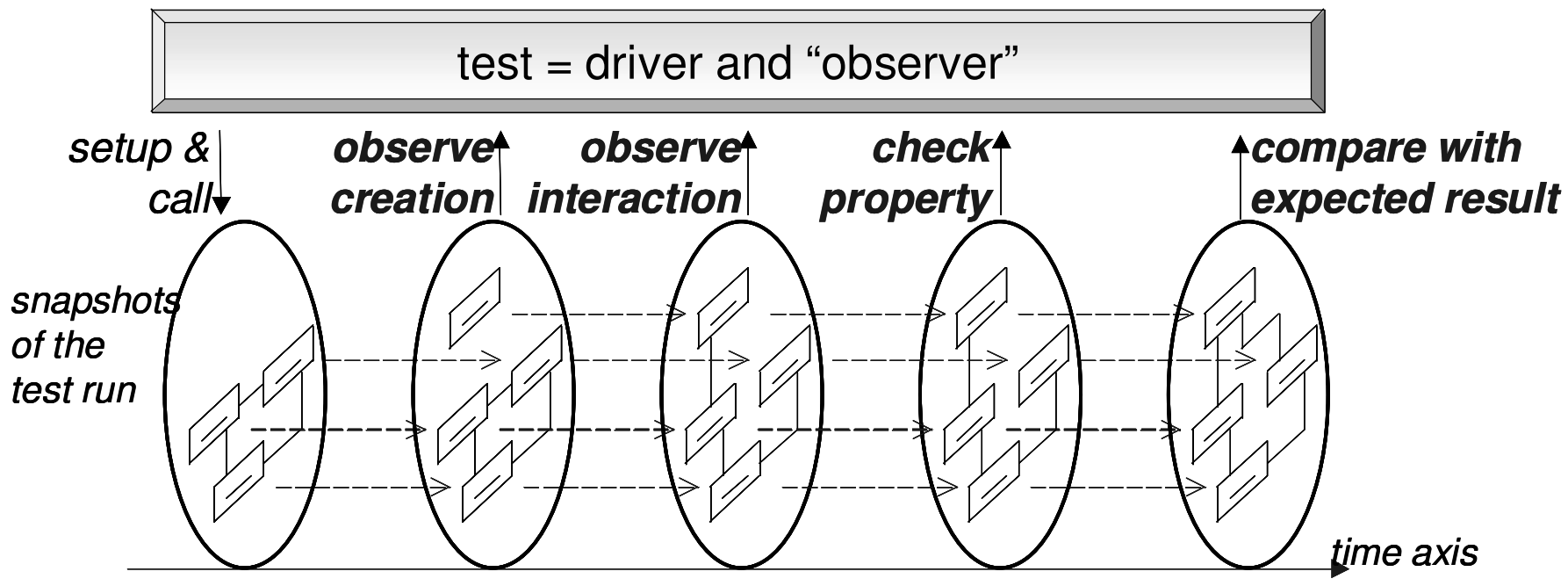

**Fig. 7.** A test case acts as observation.

Fig. 7 also shows that tests do not necessarily constrain their observation to "externally visible behavior", but can make observations on local structure, internal interactions or state properties even during the system run. Therefore, it is essential to identify, which tests are regarded as "internal" and are evolving together with the transformed system and which tests need to remain unchanged, because they describe external properties of the system. Tests in one categorization can roughly be divided into unit tests, integration tests and acceptance tests.

*Unit and integration tests* focus on small parts of the system (classes or subsystems) and usually take a deep look into system internals. It therefore isn't surprising that these kinds of tests can become erroneous after a transformation of the underlying models. Indeed, these tests are usually transformed together with the code models. For example, moving an attribute upward as shown in Fig. 6 induces object diagrams with Guest-objects to be adapted accordingly by providing a concrete value for that attribute. In this case it may even be of interest to clone tests in order to allow for different values to be tested. Contrary, tests may also become obsolete if function-ality or data structure is simplified. The task of transforming test models together with production code models can therefore not be fully automated.

Unit and integration tests are usually provided by the developer or test teams that have access to the systems internal details. Therefore, these are usually “glass box tests”. *Acceptance tests*, instead, are “black box” tests that are provided by the user (although again realized by developers) and describe external properties of the system. These tests must be a lot more robust against changes of internal structure. Fig. 8 illustrates a commuting diagram that shows how an observation remains invariant under a test.

To achieve robustness, acceptance tests should be modeled against the published interfaces of a system. In this context “published” means that parts of the system that are explicitly marked as externally visible and therefore usually rather stable. Only explicit changes of requirements lead to changes of these tests and indeed the adaptation of requirements can very well be demonstrated through adaptation of these test models followed by the transformations necessary to meet these tests afterwards in a “test-first-approach”. An adapted approach also works for changes in the interfaces between subsystems.

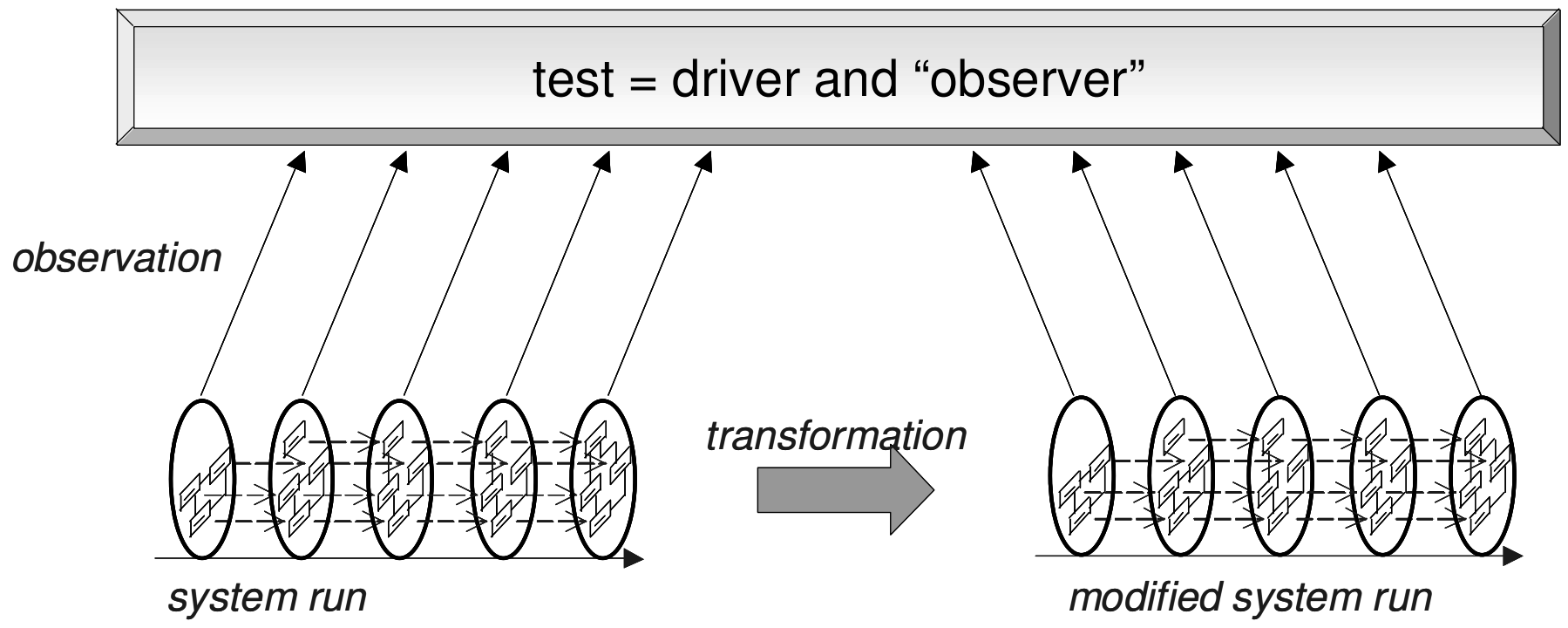

**Fig. 8.** The transformed system model is invariant under a test observation.

To increase stability of acceptance tests in transformational development, it has proven useful to follow a number of standards for test model development. These are similar to coding standards and have been found useful already before the combination with the transformational approach:

- In general an acceptance test should be abstract, by not trying to determine every detail of the tested part of the system.
- A test oracle should not try to determine every part of the output and the resulting data structure, but concentrate on important details, e.g. by ignoring uninteresting objects and attribute values.
- OCL property descriptions can often be used to model a range of possible results instead of determining one concrete result.

- Query-methods should be used instead of direct attribute access. This is more stable when the data structure is changed.
- It should not be tried to observe internal interactions during the system run. This means that sequence diagrams that are used as test drivers concentrate on triggers and on interactions with the system border only.
- Explicitly published interfaces that are regarded as highly stable should be introduced and acceptance tests should focus on these interfaces.

## 6 Conclusions

The proposal made in this paper can be summarized as a pragmatic approach to model-based software development. It uses models as primary artifact for requirements and design documentation, code generation and test case development. A transformational approach to model evolution allows an efficient adaption of the system to changing requirements and technology, optimizing architectural design and fixing bugs. To ensure the quality of such an evolving system, intensive sets of test cases are used. They are modeled in the same language, namely UML, and thus exhibit a good integration and allow to model system and tests in parallel.

However, the methodology sketched here still is a major proposal, adequate to be described in proceedings about the future of software technology. Major efforts still have to be done. On the one hand, even though initial works on various model transformations do exist, they are not very well put in context and not very well integrated with the UML in its current version. For example, it remains a challenge to provide automated support for the adaptation necessary for sequence diagrams that are affected by statechart changes. Similarly, object diagrams usually are affected when the underlying class diagrams are changed. At least in the latter case, a number of results can be reused from the area of database schema evolution. But neither the pragmatic methodology, nor theoretic underpinning are very well explored yet, even though there is currently intensive research in the area of test theory development.

On the other hand, model based evolution will become successful only if well assisted by tools. This includes parameterized code generators for the system as well as for executable test drivers, analysis tools and comfortable help for systematic transformations on models. Today, there is not yet enough progress in these direction.

As a further obstacle, these new techniques, namely an executable sublanguage of the UML as well as a lightweight methodological use of models in a development process are both a challenge to traditional software engineering. They exhibit new possibilities and problems. Using executable UML allows to program in a more abstract and efficient way. This may finally downsize projects and decrease costs. The free resources can alternatively be used within the project for additional validation activities, such as reviews, additional tests or even a verification of critical parts of the system. Techniques such as refactoring and test-first design will change software engineering and add new elements to its portfolio.

To summarize, models can and should be used as described in this paper, but of course they are not restricted to. Instead it should be possible to have a variety of sophisticated analysis and manipulation techniques available that ideally operate on the same notations. These techniques should be used whenever appropriate. Even though, data and control flow techniques, model checking or even interactive verification techniques are already available, they still have to find their broad application to models.

The Model Driven Architecture (MDA) [6] initiative from the OMG has currently received lots of interest and several tools and approaches like "executable UML" are coming up to assist this approach. However, in the MDA community there is generally a belief that MDA only works for large and rather inflexibly run projects. It may therefore remain a challenging task, to derive efficient tools as described above and to adapt the currently used development processes to this very model-centric and agile development process.

## Acknowledgements

I would like to thank Markus Pister, Bernhard Schätz und Tilman Seifert for commenting an earlier version of the paper as well as for valuable discussions. This work was partially supported by the Bayerisches Staatsministerium für Wissenschaft, Forschung und Kunst and through the Bavarian Habilitation Fellowship, the German Bundesministerium für Bildung und Forschung through the Virtual Software Engineering Competence Center (ViSEK).

## References

1. OMG - Object Management Group. Unified Modeling Language Specification. V1.5. 2002.
2. Lutz M., Ascher D. Learning Python. O'Reilly & Associates. 1999.
3. W3C. Extensible Markup Language (XML) 1.0 (2nd ed.). http://www.w3.org/xml, 2000.
4. Beck, K. Extreme Programming explained, Addison-Wesley. 1999.
5. Cockburn, A. Agile Software Development. Addison-Wesley, 2002.
6. OMG. Model Driven Architecture (MDA). Technical Report OMG Document ormsc/2001-07-01, Object Management Group, 2001.
7. Jürjens J. UMLsec: Extending UML for Secure Systems Development. In: J.-M. Jezequel, H. Hussmann, S. Cook (eds): UML 2002 - The Unified Modeling Language, pages:412-425, LNCS 2460. Springer Verlag 2002.
8. Rumpe, B. Agiles Modellieren mit der UML. Habilitation Thesis. To appear 2003.
9. Rumpe, B. Executable Modeling with UML. A Vision or a Nightmare? In: Issues & Trends of Information Technology Management in Contemporary Associations, Seattle. Idea Group Publishing, Hershey, London, pp. 697-701. 2002.
10. Siedersleben J., Denert E. Wie baut man Informationssysteme? Überlegungen zur Standardarchitektur. Informatik Spektrum, 8/2000:247-257, 2000.
11. Link J., Fröhlich P. Unit Tests mit Java. Der Test-First-Ansatz. dpunkt.verlag Heidelberg, 2002.

12. Beck K. Aim, Fire (Column on the Test-First Approach). IEEE Software, 18(5):87-89, 2001.
13. Agile Manifesto. http://www.agilemanifesto.org/. 2003.
14. Fowler M. Refactoring. Addison-Wesley. 1999.
15. Opdyke W., Johnson R. Creating Abstract Superclasses by Refactoring. Technical Report. Dept. of Computer Science, University of Illinois and AT&T Bell Laboratories. 1993
16. Philipps J., Rumpe B.. Refactoring of Programs and Specifications. In: Practical foundations of business and system specifications. H.Kilov and K.Baclawski (Eds.), 281-297, Kluwer Academic Publishers, 2003.
17. Kiczales G., Lamping J., Mendhekar A., Maeda C., Lopez C., Loingtier J.-M., Irwin J. Aspect-Oriented Programming. In ECOOP'97 - Object Oriented Programming, 11th European Conference, Jyväskylä, Finnland, LNCS 1241. Springer Verlag, 1997.
18. Ambler S. Agile Modeling. Effective Practices for Extreme Programming and the Unified Process. Wiley & Sons, New York, 2002.
19. Binder R. Testing Object-Oriented Systems. Models, Patterns, and Tools. Addison-Wesley, 1999.
20. Briand L. and Labiche Y. A UML-based Approach to System Testing. In M. Gogolla and C. Kobryn (eds): «UML» - The Unified Modeling Language, 4th Intl. Conference, pages 194-208, LNCS 2185. Springer, 2001.
21. Warmer J., Kleppe A. The Object Constraint Language. Addison-Wesley. 1998.
22. Rumpe B. E-Business Experiences with Online Auctions. In: Managing E-Commerce and Mobile Computing Technologies, Julie Mariga (Ed.) Idea Group Inc., 2003.
23. Rumpe B., Schröder A. Quantitative Survey on Extreme Programming Projects. In: Third International Conference on Extreme Programming and Flexible Processes in Software Engineering, XP2002, May 26-30, Alghero, Italy, pg. 95-100, 2002.